\documentclass[amsmath,amssymb,twocolumn,superscriptaddress,nofootinbib]{revtex4-2}

\usepackage{amsthm}
\usepackage{CJK}
\usepackage[utf8]{inputenc}
\usepackage{graphicx}
\usepackage{dcolumn}
\usepackage{bm}
\usepackage{braket}
\usepackage{color}
\usepackage{units}
\usepackage{multirow}
\usepackage{array}
\usepackage{dsfont}
\usepackage{float}
\usepackage{xcolor}
\graphicspath{{Figures/}}
\usepackage{comment}
\usepackage{mathtools} 
\usepackage{enumitem}

\newcommand{\Tr}{\operatorname{Tr}}
\newcommand{\Prob}{\operatorname{Pr}}
\newcommand{\id}{\operatorname{id}}

\newtheorem{theorem}{Theorem}
\newtheorem{lemma}{Lemma}

\usepackage{hyperref}
\usepackage{xurl} 

\hypersetup{
breaklinks=true,
colorlinks = true,
citecolor= blue,
urlcolor= blue,
linkcolor = blue
}

\begin{document}
\begin{CJK*}{GB}{}

\title{The Quantum Chernoff Divergence in Advantage Distillation for QKD and DIQKD}

\author{Mikka Stasiuk}
\affiliation{Institute for Quantum Computing and Department of Physics and Astronomy, University of Waterloo, N2L3G1 Waterloo, Ontario, Canada}

\author{Norbert L\"utkenhaus}
\affiliation{Institute for Quantum Computing and Department of Physics and Astronomy, University of Waterloo, N2L3G1 Waterloo, Ontario, Canada}

\author{Ernest Y.-Z. Tan}
\affiliation{Institute for Quantum Computing and Department of Physics and Astronomy, University of Waterloo, N2L3G1 Waterloo, Ontario, Canada}

\begin{abstract}
Device-independent quantum key distribution (DIQKD) aims to mitigate adversarial exploitation of imperfections in quantum devices, by providing an approach for secret key distillation with modest security assumptions. Advantage distillation, a two-way communication procedure in error correction, has proven effective in raising noise tolerances in both device-dependent and device-independent QKD. Previously, device-independent security proofs against IID collective attacks were developed for an advantage distillation protocol known as the repetition-code protocol, based on security conditions involving the fidelity between some states in the protocol. However, there exists a gap between the sufficient and necessary security conditions, which hinders the calculation of tight noise-tolerance bounds based on the fidelity. We close this gap by presenting an alternative proof structure that replaces the fidelity with the quantum Chernoff divergence, a distinguishability measure that arises 
in symmetric hypothesis testing.
Working in the IID collective attacks model, we derive matching sufficient and necessary conditions for the repetition-code protocol to be secure (up to a natural conjecture regarding the latter case) in terms of the quantum Chernoff divergence, hence indicating that this serves as the relevant quantity of interest for this protocol. 
Furthermore, using this security condition we obtain some improvements over previous results on the noise tolerance thresholds for DIQKD.
Our results provide insight into a fundamental question in quantum information theory regarding the circumstances under which DIQKD is possible. 
\end{abstract}

\maketitle

\end{CJK*}

\section{Introduction}
Quantum key distribution (QKD)~\cite{BB84,Eke91} is founded on the goal of extracting secret keys from measurement correlations between quantum systems. Much of QKD research is focused on ``device-dependent'' protocols, where it is assumed that devices execute quantum operations within specified tolerances. However, implementations of device-dependent QKD may suffer from attacks that leverage discrepancies between the physical devices and the models in the security proofs, undermining the security of such implementations \cite{FQT+07, GLL+11, JSK+15}. 
Device-independent (DI) QKD~\cite{PAB+09, Sca12} is an alternative that seeks to surmount this weakness by deriving security proofs with only minimal assumptions on the devices involved. This is made possible by taking advantage of quantum correlations that violate a Bell inequality~\cite{CHSH69,BCP+14}, which can allow for the production of secret keys even without characterizing the measurements performed by the devices~\cite{PAB+09, Sca12}. In effect, DIQKD lacks various vulnerabilities of device-dependent QKD that arose from requiring detailed characterization of the device measurements. 

However, the improved security prospects come at the cost of DIQKD having worse noise and loss tolerances than device-dependent QKD. 
Furthermore, there is a foundational question of more precisely characterizing the circumstances under which DIQKD is possible, as it has been shown for instance that not all correlations that violate Bell inequalities can be used for DIQKD~\cite{FBL21}.
Hence a direction of interest in current DIQKD research is to find methods that improve the noise tolerance beyond the values in existing protocols. Thus far, some methods that have been proposed include noisy pre-processing \cite{HST+20, WAP21, SBV+21,MPW22,BFF24} (where trusted noise is added to the trusted parties' raw measurement outputs), random key measurements \cite{SGP+21,MPW22} (where random measurement choices are selected for key generation), random postselection \cite{XZZ+22} (where some rounds are randomly discarded based on the measurement outcomes), and advantage distillation~\cite{TLR20,HT22} (where a two-way communication procedure is substituted into the error correction step of the protocol). Some of these methods were also previously studied in the context of device-dependent QKD~\cite{Mau93,rennerthesis,BA07,MRD+09,arx_Myh11}. Our work focuses on enhanced performance of DIQKD by means of the last technique, i.e.~advantage distillation.

Under the assumption of independent and identically distributed (IID) collective attacks, previous works~\cite{TLR20,HT22} have studied the noise tolerance of DIQKD when implementing
an advantage distillation protocol known as the repetition-code protocol. 
In particular, in~\cite{TLR20} a sufficient condition was derived for this protocol to be secure, based on the fidelity between some states that arise in the protocol. A conjectured fidelity-based necessary condition for security also exists~\cite{HT22}, but it does not match the aforementioned sufficient condition. 
This gap between the sufficient and necessary conditions presents difficulties in characterizing the noise tolerance of the protocol --- for instance, in~\cite{HT22} a semidefinite programming (SDP) algorithm was developed to compute arbitrarily tight bounds on the fidelity, but the resulting noise tolerance improvements were limited by the fact that these fidelity-based conditions do not provide an exact characterization of when the protocol is secure. 

Hence in this work, we aim to address this issue by investigating relevant quantities that could replace the role of fidelity in those security proofs, ideally yielding matching sufficient and necessary conditions. 
Specifically, an interesting candidate is the \emph{quantum Chernoff divergence}~\cite{NS09,ACM+07,ANS+08}, defined for arbitrary quantum states $\rho$ and $\sigma$ as being given by the following infimum over a parameter $s\in(0,1)$:
\begin{equation} 
    \xi(\rho, \sigma) \coloneqq -\log \left( \inf_{0 < s < 1} \Tr \left( \rho^{s} \sigma^{1-s} \right) \right).
\end{equation}
For easier comparison to the previous fidelity-based results, in this work we instead mainly present our results in terms of the argument of the logarithm in the above expression, which can be referred to as the \emph{non-logarithmic quantum Chernoff divergence} (NQCD):
\begin{equation} \label{NQCD def}
    Q(\rho, \sigma) \coloneqq \inf_{0 < s < 1} \Tr \left( \rho^{s} \sigma^{1-s} \right).
\end{equation}

The quantum Chernoff divergence arises in a quantum information task known as symmetric hypothesis testing; refer to e.g.~\cite{NS09,ACM+07} for full details. Qualitatively, this task can be described as distinguishing states of the form $\rho^{\otimes k}$ and $\sigma^{\otimes k}$. The characteristic property of the quantum Chernoff divergence is that it describes how well this task can be achieved (for large $k$). More specifically, it gives an asymptotic characterization of the trace distance $d\left(\rho^{\otimes k}, \sigma^{\otimes k} \right)$ (see e.g.~\cite{NC10} for an introduction to the trace distance and its relation to distinguishing probability), via the following relation (derived as Theorem~2.2 of~\cite{NS09} and Eq.~(1) of~\cite{ACM+07}):
\begin{equation} \label{NQCD_asymp}
-\log Q(\rho,\sigma) = \lim_{k \rightarrow \infty} \left( - \frac{1}{k}\log \frac{1-d\left(\rho^{\otimes k}, \sigma^{\otimes k} \right)}{2} \right).
\end{equation}

Our main finding in this work is that by using this fundamental property of the NQCD, we can derive security conditions for the repetition-code protocol. (This is because we show that in the repetition-code protocol, a critical part of the security analysis can be reduced to having Eve distinguish between states of the form $\rho^{\otimes k}$ and $\sigma^{\otimes k}$, which is usefully characterized by the NQCD as described above.)
Specifically, we obtain a sufficient condition (in terms of the NQCD) for device-independent security of the repetition-code protocol against IID collective attacks. Furthermore, we also derive a necessary condition for security of the protocol, which matches the sufficient condition up to a natural conjecture regarding the structure of Eve's optimal attack. Hence our results indicate that the NQCD is the central quantity of interest for studying the security of this protocol.
(We remark that while the focus of our work is on DIQKD, our results also serve to generalize the findings of~\cite{BA07} for device-dependent QKD to a broader class of device-dependent QKD protocols; we discuss this further in Sec.~\ref{sec:securitycond}.)

Using this condition, we obtain lower bounds on the noise tolerance of some DIQKD scenarios under the depolarizing-noise model, 
achieving improvements of up to $2.7\%$ compared to the fidelity-based condition~\cite{TLR20,HT22}, and an improvement of $0.23\%$ compared to the best previous noise tolerances~\cite{TSB+22} (note that the protocol in that work only uses one-way error correction, but it implements both noisy pre-processing and random key measurements).
While the latter improvement is fairly small, this arises mainly from the fact that the existing techniques to evaluate the NQCD in the device-independent setting do not yield very tight bounds. 
Since our work highlights the NQCD as the critical quantity that characterizes security of the repetition-code protocol (in the IID collective attacks setting, at least), future study of this protocol could be focused on developing better techniques to bound this quantity, which should improve the noise tolerances further.

\section{Protocol description}
We outline a general DIQKD protocol between two trusted parties. These two parties, Alice and Bob, possess $M_A$ and $M_B$ possible measurement choices, indexed by $x \in \{0, ..., M_A -1\}, \; y \in \{0, ..., M_B -1\} $. We shall often refer to the labels $x,y$ as Alice and Bob's inputs, in the sense that they can be viewed as classical inputs Alice and Bob supply to the devices in order to specify which measurement is performed. In each round, 
an untrusted source Eve sends Alice and Bob each a part of some entangled state, and retains some extension of the state for herself as side-information.
Alice and Bob then choose some inputs $x,y$ to supply to their devices, which measure the state and provide some outputs, that are recorded in registers $\hat A_x$ and $\hat B_y$ for Alice and Bob respectively.
Following the QKD protocol structure outlined in~\cite{rennerthesis}, after collecting outputs from sufficiently many rounds, Alice and Bob use some subset of their data to perform parameter estimation, 
in which they decide whether to accept or abort the protocol.
If they accept, some or all of the remaining rounds are then used to generate the final key, using some classical post-processing steps we describe below.
We shall focus only on protocols where all measurements are binary-outcome and there is only a single pair of measurements used for key generation, which we shall take to correspond to the labels $x=y=0$. 

We perform the security analysis in the IID collective attacks setting, where it is assumed that all states and measurements are IID. In this case, it is enough to consider Alice, Bob and Eve's single round composite state, written as $\sigma_{ABE}$, where $E$ denotes Eve's purification of Alice and Bob's states. 
Also, we shall denote the hermitian operators describing the measurements for inputs $x,y$ as $A_x,B_y$ respectively (not to be confused with the registers $\hat A_x, \hat B_y$ that store the outputs).
When the key-generating measurements are performed, the resulting state produced between the three parties is a classical-classical-quantum state of the form
\begin{equation}\label{ccq_after_key_gen}
    \sum_{ab} \Prob(\hat A_0 = a, \hat B_0 = b) \ket{ab} \!\bra{ab}_{\hat A_0 \hat B_0} \otimes \sigma_{E|ab},
\end{equation}
where $\sigma_{E|ab}$ is Eve's single-round state conditioned on Alice and Bob performing the key-generating measurements and getting output values $\hat A_0=a, \hat B_0 = b$.

To simplify the security proof, we also impose a symmetrization step\footnote{For the protocol we consider, this symmetrization step can be omitted in practice without significantly affecting the sufficient conditions for security, but it may affect the necessary conditions --- we elaborate on this point in Sec.~\ref{sec:discussion}.}, which involves Alice publicly communicating to Bob a uniformly random bit $T$, followed by having each of them XOR their output value with $T$. Let $\widetilde A_x$ and $\widetilde B_y$. denote Alice and Bob's registers respectively that store the symmetrized values (for inputs $x$ and $y$), i.e.~so we have $\widetilde A_x = \hat A_x \oplus T$ and $ \widetilde B_y = \hat B_y \oplus T$. 
We shall use $\Prob(ab|xy)$ to denote the probability distribution on registers $\widetilde A_x \widetilde B_y$, i.e.~the distribution of output values \emph{after} symmetrization, conditioned on the input pair $x, y$ being used. 

With this notation, when the key-generating measurements are performed, the composite state between the three parties after symmetrization is
\begin{equation} \label{ccq_after_symm}
    \sum_{ab} \Prob(ab|00) \ket{ab} \!\bra{ab}_{\widetilde A_0 \widetilde B_0} \otimes \rho_{ET|ab},
\end{equation}
where $\rho_{ET|ab}$ denotes Eve's single-round state conditioned on $\widetilde A_0 = a , \widetilde B_0 = b$, and is given by 
\begin{equation}\label{eve_single_round}
    \rho_{ET|ab} = \frac{1}{2} \sigma_{E|ab} \otimes \ket{0}\!\bra{0}_T + \frac{1}{2} \sigma_{E|\overline a \overline b} \otimes \ket{1}\!\bra{1}_T,
\end{equation}
where $\overline a$ denotes $a \oplus 1$ and similarly for $\overline b$.
The symmetrization step enables the outcome probabilities $\Prob(ab|00)$ for the key-generating measurements to be written in terms of a single parameter: the quantum bit error rate (QBER) $\epsilon$, which we define in this context to be the probability that $\widetilde A_0 \neq \widetilde B_0$. Explicitly, we have $\text{Pr}(01|00)=\text{Pr}(10|00)=\frac{\epsilon}{2}$ and $\text{Pr}(00|00)=\text{Pr}(11|00)=\frac{1-\epsilon}{2}$. 

We focus on deriving the conditions for security in the asymptotic limit of infinitely many rounds. In this setting, we can make the simplifying assumption (see e.g.~\cite{rennerthesis,GLL21} for further discussion) that the condition for the parameter-estimation step to accept is that the statistical estimates for the probability distribution $\Prob(ab|xy)$ exactly match the values for an honest implementation of the protocol. 
Under the IID assumption, these estimates are arbitrarily accurate given enough rounds, which implies (by the arguments in~\cite{rennerthesis,GLL21}) that for the asymptotic analysis we can focus only on states producing the same distribution $\Prob(ab|xy)$ as in an honest implementation.

We now present the repetition-code protocol for advantage distillation, following the description in e.g.~\cite{rennerthesis}. 
First, Alice and Bob gather the outputs from their devices over some number of key-generating rounds, which we shall denote as $N$.
They split the $N$ rounds into blocks of $k$ rounds and execute the following procedure on each individual block. Focusing on one block, let $\widetilde{\mathbf{A}}_0$ and  $\widetilde{\mathbf{B}}_0$ denote Alice and Bob's (symmetrized) values respectively for this block. For this block, Alice independently generates a uniform random bit $C \in \{0,1\}$ and sends $\mathbf{M} = \widetilde{\mathbf{A}}_0 \oplus (C,...,C)$ to Bob, who computes $\widetilde{\mathbf{B}}_0 \oplus \mathbf{M}$. If $\widetilde{\mathbf{B}}_0 \oplus \mathbf{M} = (C', ..., C')$ for some bit $C' \in \{0, 1\}$, then Bob accepts the block and sends a bit $D=1$ to Alice. Otherwise, he rejects the block and sends $D=0$. (Note that this accept/reject value is for individual blocks, not the overall protocol.)
After performing this procedure across all the blocks, Alice and Bob respectively keep the strings of $C$ and $C'$ values from the accepted blocks. Alice and Bob then perform a one-way error-correction procedure on these strings, and implement privacy amplification to produce their final key (see~\cite{rennerthesis} for details). 

Let $\mathbf{ET}$ denote Eve's side information (including the symmetrization bits) across all rounds of one block, and let $H(\cdot)$ denote the von Neumann entropy.
It was shown in~\cite{rennerthesis,DW05} that
if the values $C,C'$ produced in each block satisfy the Devetak--Winter condition
\begin{equation}\label{eq:DWcondition}
    H(C|\mathbf{ETM};D=1)-H(C|C';D=1)>0,
\end{equation}
then the above procedure achieves a positive secret key rate in the asymptotic limit of infinitely many rounds $N$. Conversely, the results in~\cite{Mau93} for binary symmetric channels imply (see also the discussion in~\cite{BA07}) that if in each block, Eve can use her registers $\mathbf{ETM}$ to construct a guess $C''$ for the bit $C$ that satisfies
\begin{align}\label{guessC''}
\Prob(C \neq C'' | D=1) \leq \Prob(C \neq C'| D=1),
\end{align}
then the repetition-code protocol as described above\footnote{Note that in particular, this means Alice and Bob perform the one-way error-correction procedure on the strings of $C,C'$ bits ``directly'', without further processing them first. We leave for future work the question of whether such processing (for instance, possibly applying the block-processing procedure iteratively) could further improve the noise tolerance.} cannot achieve a positive asymptotic secret key rate. 
Hence in our work, we shall focus on studying conditions under which~\eqref{eq:DWcondition} holds (implying positive asymptotic keyrate is achievable) or~\eqref{guessC''} holds (implying positive asymptotic keyrate is not achievable).

\section{Results}
In this section, we first present the sufficient and necessary conditions for security as Theorems~\ref{theorem:1} and~\ref{theorem:2} respectively. After that, we compute the improvement in noise tolerance with these new findings, and compare our results to noise tolerances computed for one-way protocols as well as the previous fidelity-based security condition. 

\subsection{Security conditions}
\label{sec:securitycond}

Our first result is a sufficient condition for the repetition-code protocol to be secure (the proof of this theorem is given in Appendix~\ref{app:sufproof}, based on an approach similar to~\cite{BA07,TLR20}):
\begin{theorem}\label{theorem:1}
For the DIQKD repetition-code protocol described above, if Eve's single-round conditional states 
as described in~\eqref{ccq_after_symm}--\eqref{eve_single_round} satisfy
\begin{align}\label{eq:suffQ}
    Q(\rho_{ET|00}, \rho_{ET|11}) > \frac{\epsilon}{1-\epsilon},
\end{align}
then there exists some value $k_\mathrm{min}$ such that protocol achieves positive asymptotic keyrate as long as the block size $k$ is at least $k_\mathrm{min}$.
\end{theorem}

We note that there is a subtle limitation in the above statement. Specifically, with our proof approach, we currently do not have an explicit value for the minimum block size $k_\mathrm{min}$ in the theorem statement, unlike the previous fidelity-based results in~\cite{TLR20}. 
This is because our proof relies on a particular function $g(k)$ (see Appendix~\ref{app:sufproof}) which is known to converge to $0$ in the large-$k$ limit~\cite{NS09,ACM+07,ANS+08}; however, the current known convergence bounds~\cite{AMV12} for this function have a dependence on the dimension and minimal nonzero eigenvalues of the states $\rho_{ET|00},\rho_{ET|11}$. In the context of DIQKD, we do not know the dimensions or eigenvalues of these states, which prevents us from finding an explicit value for $k_\mathrm{min}$. Doing so would require a modification of the bounds in~\cite{AMV12} to remove their dependence on these parameters; we leave this for future work.

While the full proof of Theorem~\ref{theorem:1} is deferred to Appendix~\ref{app:sufproof}, we present here an informal outline of the main ideas:
\begin{proof}[Informal proof outline:]
Observe that conditioned on $D=1$ and the public message taking some value $\mathbf{M} = \mathbf{m}$, Eve's side-information state for the four possible values $(0,0), \, (0, 1), \, (1, 0)$ and $(1, 1)$ for $(C, C')$ takes the form $\rho_{\mathbf{ET}|\mathbf{mm}}, \, \rho_{\mathbf{ET}|\mathbf{m \overline m}}, \, \rho_{\mathbf{ET}|\mathbf{\overline mm}}$ and $\rho_{\mathbf{ET}|\mathbf{\overline m \overline m}}$ respectively, where $\mathbf{\overline m}$ denotes the bitflip of $\mathbf{m}$, and 
\begin{equation}\label{eq:defnrhoETmm}
\rho_{\mathbf{ET}|\mathbf{mm}} = \rho_{ET|m_1 m_1}\otimes ... \otimes \rho_{ET|m_k m_k},
\end{equation}
where the states $\rho_{ET|m_j m_j}$ refer to the single-round states as defined in the formula~\eqref{eve_single_round}. 

We now need to quantify Eve's uncertainty about 
the value $C$, in the sense of lower-bounding $H(C|\mathbf{ETM};D=1)$ in the condition~\eqref{eq:DWcondition}.
To do so, we first informally note that the cases where $C \neq C'$ are rare for large block sizes\footnote{The case where $C \neq C'$ corresponds to scenarios where every round in Alice's block is flipped during key distillation. For large block sizes, this happens far less often.}, which lets us remove them using a suitable continuity bound (in the full proof in Appendix~\ref{app:sufproof}, this is formalized via the bound~\eqref{eq:contbnd}). This allows us to focus mainly on Eve's information when $C=C'$. Conditioned on this, the classical-quantum state (with Eve's side-information) for each accepted block takes the following form:
\begin{equation}\label{CET_state}
    \widetilde \rho_{C\mathbf{ET}} = \frac{1}{2} \ket{0} \! \bra{0}_C \otimes \rho_{\mathbf{ET}|\mathbf{mm}} + \frac{1}{2}\ket{1} \! \bra{1}_C  \otimes \rho_{\mathbf{ET}| \mathbf{\overline m \overline m}}.
\end{equation}
(Note that $\widetilde \rho_{C\mathbf{ET}}$ as defined above has a dependence on the message $\mathbf{m}$, which we will not explicitly write in the notation for the state.)
This is the state we focus on in the main part of the proof.
Conveniently, it has the following property:
\begin{lemma}\label{lemma:1}
Consider the state $\widetilde \rho_{C\mathbf{ET}}$ as defined in~\eqref{CET_state}, for some block size $k$. Let us define the following state:
\begin{equation}\label{special_case_CETstate}
    \overline \rho_{C\mathbf{ET}} = \frac{1}{2}\ket{0} \! \bra{0}_C  \otimes \rho_{ET|00}^{\otimes k} + \frac{1}{2}\ket{1} \! \bra{1}_C  \otimes \rho_{ET|11}^{\otimes k},
\end{equation}
where $\rho_{ET|00},\rho_{ET|11}$ are Eve's single-round conditional states as described in~\eqref{ccq_after_symm}--\eqref{eve_single_round}.
Then 
\begin{equation}\label{equal_cond_entropies}
    H(C|\mathbf{ET})_{\widetilde \rho} = H(C|\mathbf{ET})_{\overline \rho}.
\end{equation}
\end{lemma}

The point of this lemma is that it allows us to focus on simplified states of the form in~(\ref{special_case_CETstate}), since they have the same entropies $H(C|\mathbf{ET})$ as those of the form~\eqref{CET_state}. 
In particular, observe that conditioned on each value of $C$, the side-information registers $\mathbf{ET}$ have a convenient tensor-power form --- this is what allows us to later reduce the analysis to the NQCD, by using the characteristic property~\eqref{NQCD_asymp}.
(As an alternative to this lemma, we remark that the state~(\ref{special_case_CETstate}) could also be viewed as arising naturally in a modified version of the repetition-code protocol; we describe some details of this in Appendix~\ref{app:modified}.) We prove Lemma~\ref{lemma:1} in Appendix~\ref{app:Uinvariance} by exploiting the unitary invariance of the von Neumann entropy: basically, we observe that there is a unitary $U_X$ that transforms $\rho_{ET|ii}$ to $\rho_{ET|\overline i \overline i}$. For a given message $\mathbf{m}$, we can use this to construct a unitary
acting on $\mathbf{ET}$
that transforms the state $\widetilde \rho_{C\mathbf{ET}}$ to $\overline \rho_{C\mathbf{ET}}$, which gives the desired result.

The remaining task in the proof to essentially to find a lower bound on $H(C|\mathbf{ET})_{\overline \rho}$, by using the specific structure of $\overline \rho_{C\mathbf{ET}}$.
The approach we use for this is similar to the proof in \cite{TLR20}, but instead of using a fidelity-based lower bound on the conditional entropy $H(C|\mathbf{ET})_{\overline \rho}$, we use one that involves the NQCD between $\rho_{ET|00}$ and $\rho_{ET|11}$. This bound on $H(C|\mathbf{ET})_{\overline \rho}$ in terms of $Q(\rho_{ET|00}, \rho_{ET|11})$ allows us to prove that whenever $Q(\rho_{ET|00}, \rho_{ET|11})$ satisfies the theorem condition~\eqref{eq:suffQ}, the entropy $H(C|\mathbf{ETM};D=1)$ must be larger than $H(C|C'; D=1)$ at sufficiently large block size $k$, hence implying positive asymptotic keyrate is achievable (by the condition~\eqref{eq:DWcondition}).
\end{proof}

We now turn to the necessary conditions for security, which we shall also present in terms of $Q(\rho_{ET|00}, \rho_{ET|11})$. 
Specifically, we prove the following almost-complementary result to Theorem~\ref{theorem:1} (with the proof given in Appendix~\ref{app:necproof}):
\begin{theorem}\label{theorem:2}
For the DIQKD repetition-code protocol described above, if Eve's single-round conditional states as described in~\eqref{ccq_after_symm}--\eqref{eve_single_round} satisfy $\rho_{ET|01}=\rho_{ET|10}$ and
\begin{equation}\label{eq:neccQ}
 Q(\rho_{ET|00}, \rho_{ET|11}) \leq \frac{\epsilon}{1-\epsilon},
\end{equation}
then  regardless of block size $k$,
the protocol does not achieve a positive asymptotic keyrate.
\end{theorem} 
Observe that Theorems~\ref{theorem:1} and~\ref{theorem:2} yield almost complementary sufficient and necessary conditions for security of the repetition-code protocol, except for the extra condition $\rho_{ET|01}=\rho_{ET|10}$ in the latter. As a speculative consideration, if we were to conjecture that Eve's states can be taken to satisfy this condition without loss of generality, then the conditions would indeed become exactly complementary. We remark that it does seem plausible to conjecture that Eve can indeed always perform an operation on her side-information such that this condition is fulfilled (while keeping the states $\rho_{ET|00},\rho_{ET|11}$ unchanged), because it essentially corresponds to having Eve ``erase'' her side-information whenever Alice and Bob did not get the same output value. In fact, this property indeed holds for the device-dependent QKD protocols considered in~\cite{BA07}, and was used in their security proof. However, it does not straightforwardly generalize to DIQKD (as discussed in~\cite{HT22}), and hence proving this conjecture in this setting would require further work.

Again, while we defer the full proof of Theorem~\ref{theorem:2} to Appendix~\ref{app:necproof}, we informally outline some main ideas here:
\begin{proof}[Informal proof outline:]
Our approach is similar to the proof of Proposition 3 in \cite{HT22}, but again with the relevant quantity being the NQCD rather than the fidelity. We begin the same observations as in the Theorem~\ref{theorem:1} proof, considering the four possible values for $(C,C')$ and the corresponding side-information states~\eqref{eq:defnrhoETmm}. However, one difficulty encountered here as compared to the Theorem~\ref{theorem:1} proof is that the continuity argument we previously used to remove the $C \neq C'$ cases does not seem sharp enough to give a nontrivial result in this case. 
This is essentially the reason for the additional condition in Theorem~\ref{theorem:2} that Eve's states satisfy $\rho_{ET|01}=\rho_{ET|10}$; this condition ends up allowing us to remove the $C \neq C'$ cases from the analysis (the same approach was used in~\cite{HT22}).

With those cases removed, the proof again reduces to studying states of the form~\eqref{CET_state}. By following a similar argument as in Lemma~\ref{lemma:1}, we also reduce the analysis to states of the form~\eqref{special_case_CETstate}. We then exploit the tensor-power structure of Eve's side-information states in~\eqref{special_case_CETstate}, to show that Eve can produce a guess $C''$ for $C$ such that the probability of her guess being wrong is upper bounded in terms of $Q(\rho_{ET|00}, \rho_{ET|11})$. With this, we show that whenever the condition~\eqref{eq:neccQ} holds, the guess $C''$ satisfies the condition~\eqref{guessC''} (regardless of block size $k$), implying that positive asymptotic keyrate cannot be achieved.
\end{proof}

We highlight that while we have presented these theorems for the DI setting, they are in fact valid as security conditions for the repetition-code protocol in the device-dependent setting as well, because the proofs for these theorems by themselves do not make use of any properties specific to the DI setting. In particular, in that context our results can be viewed as an extension of the results in~\cite{BA07} for device-dependent qubit protocols --- for such protocols, our above theorems in fact reduce to precisely the same security conditions as those derived in~\cite{BA07}. However, our results are more general in that they also apply for this repetition-code procedure performed using higher-dimensional qudits, as long as the measurements only have two outcomes. (While qudit protocols were also investigated in~\cite{BA07}, a different form of advantage distillation procedure was considered for those protocols, hence those results are not directly comparable to ours.)

As a comparison, we now state the results derived in previous works (for the DI setting) that were instead based on the fidelity $F(\rho_{ET|00}, \rho_{ET|11})$. First, Theorem~1 of~\cite{TLR20} states that a sufficient condition for the protocol to achieve positive asymptotic keyrate (for large enough $k$) is
\begin{align}\label{eq:suffF}
    F(\rho_{ET|00}, \rho_{ET|11})^2 > \frac{\epsilon}{1-\epsilon}.
\end{align}
Furthermore, Proposition~3 of~\cite{HT22} states\footnote{In fact, Proposition~3 of~\cite{HT22} also covers a variant of the fidelity known as the ``pretty-good fidelity''~\cite{LZ04,Aud14comp,IRS17}, but our subsequent discussions regarding the condition~\eqref{eq:neccF} are also valid for the pretty-good fidelity, as it is also 
lower bounded by $Q(\rho_{ET|00}, \rho_{ET|11})$~\cite{Aud14comp}.
} that if Eve's conditional states satisfy $\rho_{ET|01}=\rho_{ET|10}$ and
 \begin{equation}\label{eq:neccF}
     F(\rho_{ET|00}, \rho_{ET|11}) \leq \frac{\epsilon}{1-\epsilon},
 \end{equation}
then the protocol cannot achieve positive asymptotic keyrate (for any $k$).

Note that unlike our result, these conditions are not perfectly ``complementary'' to each other, as the fidelity term is squared in one but not the other. Furthermore, note that whenever the fidelity-based sufficient condition~\eqref{eq:suffF} holds, our sufficient condition~\eqref{eq:suffQ} automatically holds as well, due to the inequality $Q(\rho_{ET|00}, \rho_{ET|11}) \geq F(\rho_{ET|00}, \rho_{ET|11})^2$~(Eq.~(32) in~\cite{ANS+08}). Similarly, whenever the fidelity-based condition~\eqref{eq:neccF} holds, our condition~\eqref{eq:neccQ} automatically holds as well, by the inequality $Q(\rho_{ET|00}, \rho_{ET|11}) \leq F(\rho_{ET|00}, \rho_{ET|11})$~(Eq.~(28) in~\cite{ANS+08}). Hence any conclusion that could have been drawn about the security of the repetition-code protocol using the conditions~\eqref{eq:suffF} and~\eqref{eq:neccF} could also have been drawn using our conditions~\eqref{eq:suffQ} and~\eqref{eq:neccQ}, but not necessarily vice versa. In this sense, our results serve as an improvement over the fidelity-based conditions.

\subsection{Resulting noise tolerances}
\label{sec:noisetol}

We now use our results to compute the noise tolerances for 2-output distributions for a varying number of measurement settings, using the depolarizing noise model. Explicitly, this means that $\Prob(ab|xy)$ (recall that this denotes the distribution after symmetrization) is of the following form:
\begin{equation}\label{bipartite_correlation}
\Prob(ab|xy) = (1-2q)\Prob_{\mathrm{target}}(ab|xy)+\frac{q}{2},
\end{equation}
where $q \in [0, \frac{1}{2}]$ is the depolarizing noise parameter, and $\Pr_\mathrm{target}$ is some target distribution (after symmetrization) described by the relevant measurement choice. We now present the three scenarios we consider: in all cases,  the target distributions are generated by performing the specified measurements on the maximally entangled Bell state $\ket{\Phi^+} = \frac{1}{\sqrt{2}}(\ket{00} + \ket{11})$ (in the following, $X$ and $Z$ denote the Pauli $X$ and $Z$ observables respectively).
\begin{enumerate}[leftmargin=*]
    \item \label{case:2222} 
    Alice and Bob have two measurement settings each, described by
    \begin{align}\label{eq:2in2out}
        A_0 = Z, \; A_1 = X, \; B_0 = \frac{X+Z}{\sqrt{2}}, \; B_1 = \frac{X-Z}{\sqrt{2}}.
    \end{align}
    This corresponds to choosing the measurements such that they maximize the violation of the CHSH inequality~\cite{CHSH69} on $\ket{\Phi^+}$. These measurements have the drawback that even in the absence of noise, we have $\epsilon \neq 0$ because the key-generating measurements $A_0,B_0$ do not produce perfectly correlated outputs.
    \item \label{case:2322} 
    Alice and Bob have two and three measurement settings respectively, described by
    \begin{equation}
    \begin{gathered}
        A_0 = B_0 = Z, \\
        A_1 = X, \; B_1 = \frac{X+Z}{\sqrt{2}}, \; B_2 = \frac{X-Z}{\sqrt{2}}.
    \end{gathered}
    \end{equation}
    With this, the key-generating measurements $A_0,B_0$ achieve $\epsilon=0$ in the absence of noise, while the measurements $A_0,A_1,B_1,B_2$ are equivalent to case~\ref{case:2222} up to relabelling. This scenario was studied in the basic DIQKD protocol proposed in~\cite{PAB+09}, as well as many subsequent works~\cite{HST+20, WAP21, SBV+21,ADF+18} (sometimes with slight modifications). 
    \item \label{case:4422} 
    Alice and Bob each have 4 measurement settings, described by 
    \begin{align}
        A_0 = Z, \; A_1 = \frac{X + Z}{\sqrt{2}}, \; A_2 = X, \; A_3 = \frac{X-Z}{\sqrt{2}},
    \end{align}
    and $B_j=A_j$ for all $j\in\{0,1,2,3\}$.
    This case can be viewed as a combination of the Mayers-Yao self-test~\cite{MY98} with the measurements in case~\ref{case:2222}. Alternatively, it can be viewed as having Alice and Bob both perform all the measurements listed in case~\ref{case:2222}.
\end{enumerate}

To compute the maximum tolerable noise for the new security condition in Theorem~\ref{theorem:2}, it is necessary to establish a lower bound on $Q(\rho_{ET|00}, \rho_{ET|11})$ in terms of  
some quantity which we can bound in the device-independent context.
One such convenient quantity is the trace distance $d(\rho_{ET|00}, \rho_{ET|11})$, which is related to $Q(\rho_{ET|00}, \rho_{ET|11})$ by the following inequality (Eq.~(9) in~\cite{ACM+07}), similar to the Fuchs--van de Graaf inequality:\footnote{We could view the inequality~\eqref{eq:QlowerFvdG} as implying that as a corollary of Theorem~\ref{theorem:1}, we have that a sufficient condition for the repetition-code protocol to be secure (given sufficiently large $k$) is to have $1 - d(\rho_{ET|00}, \rho_{ET|11}) > \frac{\epsilon}{1-\epsilon}$. Since we use the inequality~\eqref{eq:QlowerFvdG} to obtain our subsequent numerical results, we can view those numerical results as being those that would be given by this corollary.
However, we present Theorem~\ref{theorem:1} in its stated form since it is a more general statement.} 
\begin{equation}\label{eq:QlowerFvdG}
    Q(\rho_{ET|00}, \rho_{ET|11}) 
    \geq
    1 - d(\rho_{ET|00}, \rho_{ET|11}).
\end{equation}
Importantly, as noted in~\cite{TLR20}, there exists a technique to compute upper bounds on $d(\rho_{ET|00}, \rho_{ET|11})$ for DI protocols. Namely, by the operational interpretation of trace distance (see e.g.~\cite{NC10}), it can be written as
$d(\rho_{ET|00}, \rho_{ET|11}) = 2 P_g(\rho_{ET|00}, \rho_{ET|11}) - 1$,
where $P_g(\rho_{ET|00}, \rho_{ET|11})$ is the distinguishing probability between the states $\rho_{ET|00}, \rho_{ET|11}$. This probability can equivalently be viewed as 
Eve's probability of guessing Alice and Bob's single-round measurement outcomes (using her side-information $ET$) conditioned on Alice and Bob getting the same outcomes, and an SDP technique for securely bounding such probabilities in device-independent protocols was developed in~\cite{TTB+16}. 
Hence to find the maximum noise value for which we can prove the protocol is secure, our approach is to apply the technique in~\cite{TTB+16} to bound the right-hand-side of~(\ref{eq:QlowerFvdG}) as a function of $q$, and find the intersection of the resulting curve with the right-hand-side of~(\ref{eq:suffQ}).

With this, we obtained the noise tolerances shown below for the three scenarios. As some points of comparison, we also include some results obtained in previous works, which we describe in more detail below.
\def\arraystretch{1.5} 
\begin{center}
\begin{tabular}{l c}
Case~\ref{case:2222}: & $7.71\%$ \\
Case~\ref{case:2322}: & $8.18\%$ \\
Case~\ref{case:4422}: & $9.56\%$ \\
Basic one-way protocol in~\cite{PAB+09}: & $7.14\%$ \\
Previous highest threshold~\cite{TSB+22}: & $9.33\%$ \\
Upper bound from~\cite{FBL21}: & $13.69\%$ 
\end{tabular}
\end{center}
From the above list, we see that case~\ref{case:4422} gives the best noise tolerances out of the cases we have considered. Regarding the other results we presented for comparison, the value of $7.14\%$ is the noise tolerance of the basic one-way protocol proposed in an early DIQKD work~\cite{PAB+09}.
As for the value of $9.33\%$, this was the noise tolerance obtained in~\cite{TSB+22}, which is the highest value achieved thus far amongst all previous works --- we highlight that our best result (case~\ref{case:4422}) surpasses this value, albeit only by a small amount. Interestingly, the protocol studied in that work only involves one-way error correction, but it achieved high noise tolerances by using the noisy pre-processing and random key measurement techniques in tandem. 
Finally, the last value we listed above is an \emph{upper} bound derived in~\cite{FBL21} on the noise tolerances achievable by a wide family of DIQKD protocols, including the advantage distillation protocol we study here (for any of the three cases). We see that our results are still quite far from saturating this upper bound (since our results show that a noise tolerance of $9.56\%$ is achievable, whereas the upper bound is $13.69\%$), so it remains unclear how tight this upper bound is. 

To more specifically compare the three cases we considered to the works~\cite{PAB+09} and~\cite{TSB+22}, we note that the analysis in the former was based on the same measurement settings as in case~\ref{case:2322} here,
and the latter is also based on a similar scenario, except with one additional measurement setting for Bob. 
Hence the scenario we considered here that is most similar to those works is essentially case~\ref{case:2322}, in which our result outperforms the noise tolerance obtained in~\cite{PAB+09} but not that in~\cite{TSB+22}.
In principle, for a fairer comparison to our results, one could analyze the noise tolerances of the protocols in those works when considering the measurement settings in case~\ref{case:4422} (which was the scenario giving the best noise tolerances in our work). 
However, this scenario was not analyzed in those works, because their proof approaches relied on a specialized reduction to qubit states that does not seem easy to apply for case~\ref{case:4422}. Hence this would not be a straightforward question to resolve, and we leave it for future work. 

It may also be of interest to compare our results to the previous works~\cite{TLR20,HT22} based on the fidelity-based security condition~(\ref{eq:suffF}), which did explore all three cases considered here. We first briefly review those previous approaches: the fidelity-based condition~(\ref{eq:suffF}) was first presented in \cite{TLR20}, where noise tolerances were computed using the Fuchs--van de Graaf inequality: 
\begin{equation}\label{eq:lowerFvdG}
F(\rho_{ET|00}, \rho_{ET|11}) \geq 
1-d(\rho_{ET|00}, \rho_{ET|11}),
\end{equation}
with $d(\rho_{ET|00}, \rho_{ET|11})$ being bounded using the SDP technique in~\cite{TTB+16}.
Subsequently, the work \cite{HT22} improved on this approach by developing an SDP algorithm to directly compute arbitrarily tight lower bounds on $F(\rho_{ET|00}, \rho_{ET|11})$, resulting in increased noise tolerance as compared to using the indirect bound~(\ref{eq:lowerFvdG}). 
We note, however, that for 2-input 2-output protocols in particular, a different sufficient security condition 
was also derived in \cite{TLR20} (see Corollary 1 in that work), based on the trace distance directly:
\begin{align}\label{eq:suffD}
    1-d(\rho_{ET|00}, \rho_{ET|11}) > \frac{\epsilon}{1-\epsilon}.
\end{align}
For such protocols, the noise tolerances found in~\cite{TLR20} via this condition (using the SDP technique of~\cite{TTB+16} to bound $d(\rho_{ET|00}, \rho_{ET|11})$) turned out to be better than those obtained by applying the fidelity-based condition~(\ref{eq:suffF}) with the SDP algorithm of~\cite{HT22}.

In light of the above points, we see that for case~\ref{case:2222} (which is a 2-input 2-output protocol), our approach here can only yield the same noise tolerance as that obtained in~\cite{TLR20} using the condition~(\ref{eq:suffD}). This is because we only used the bound~(\ref{eq:QlowerFvdG}) to check whether our sufficient security condition~(\ref{eq:suffQ}) is satisfied, in which case this becomes equivalent to checking the condition~(\ref{eq:suffD}). 
Therefore, the noise tolerance value of $q \approx 7.71\%$ we obtained for this case is the same as that obtained in~\cite{TLR20}.
(Alternatively, if the measurement settings for this case are modified to optimize the noise tolerance instead of the CHSH value, the noise tolerance can be improved to $q \approx 9.1\%$ as noted in~\cite{TLR20}.)

In the remaining cases, however, we can anticipate an improvement in our results compared to~\cite{TLR20}, recalling that it only used the inequality~(\ref{eq:lowerFvdG}) (which is of basically the same form as the inequality~(\ref{eq:QlowerFvdG}) we used here) to check whether the fidelity-based condition~(\ref{eq:suffF}) is satisfied.
On the other hand, the approach in~\cite{HT22} was to bound the fidelity directly and check whether the condition~(\ref{eq:suffF}) is satisfied, and hence it is not \textit{a priori} clear whether our approach would be able to outperform it, since we have only bounded $Q(\rho_{ET|00}, \rho_{ET|11})$ indirectly using~(\ref{eq:QlowerFvdG}). Still, it turned out in the end that our results did indeed yield an improvement over the latter. Explicitly, for case~\ref{case:2322} our noise tolerance value of $q \approx 8.18\%$ is an improvement of  
$2.1\%$ and $1.1\%$ over the results in~\cite{TLR20} and~\cite{HT22},
respectively. 
As for case~\ref{case:4422}, our noise tolerance value of $q \approx 9.56\%$ is an increase of 
$2.7\%$ and $1.3\%$ over the results in~\cite{TLR20} and~\cite{HT22},
respectively. 

\section{Discussion and conclusion} 
\label{sec:discussion}
In this work, we have shown that the NQCD can be used to obtain a sufficient condition for security of the repetition-code advantage distillation protocol in DIQKD, for the scenario of IID collective attacks. Furthermore, it also gives a precisely complementary necessary condition for security, up to a conjecture regarding the structure of Eve's attack. 
This strongly indicates that the NQCD should be the central quantity to consider in analyzing the security of this protocol, allowing us to focus our attention on this quantity in subsequent study of this topic. In particular, resolving that conjecture would imply that when considering the question of whether this protocol allows secret key generation (against IID collective attacks), it suffices to study the value of $Q(\rho_{ET|00}, \rho_{ET|11})$.

Regarding the explicit noise tolerance thresholds given by this security condition, in this work we did not find an approach to bound $Q(\rho_{ET|00}, \rho_{ET|11})$ directly in the DI setting, and instead bounded it indirectly using the trace distance, which is not necessarily optimal. However, our results in Sec.~\ref{sec:noisetol} show that even with this limitation, our approach already suffices to certify that advantage distillation can outperform one-way protocols (under the assumption of IID collective attacks), even when using the noisy pre-processing and random key measurement techniques.

Still, there is a significant gap between the results we obtained and the upper bounds on noise tolerance derived in~\cite{FBL21},
and hence there still remains potential for further improvement. 
In particular, if tighter lower bounds on $Q(\rho_{ET|00}, \rho_{ET|11})$ were established, 
we would be able to further increase our noise tolerance thresholds.
To this end, one could aim to develop
a numerical algorithm that can compute arbitrarily tight lower bounds on $Q(\rho_{ET|00}, \rho_{ET|11})$ in the DI setting, similar to the approach developed in~\cite{HT22} for bounding the fidelity.
However, such an algorithm would face the difficulty that it is not currently known whether there exists a measurement on Eve's registers that leaves $Q(\rho_{ET|00}, \rho_{ET|11})$ invariant --- this property, which does hold for the fidelity, was essential to the approach in \cite{HT22}. Nevertheless, the development of a numerical algorithm that calculates the equivalent for $Q(\rho_{ET|00}, \rho_{ET|11})$ remains an avenue for further study, which could potentially improve the noise tolerance thresholds. 

We note that apart from the depolarizing noise model, another noise model that has been studied for DIQKD is limited detection efficiency~\cite{Ebe93}, for instance in~\cite{PAB+09,ML12,TLR20,HST+20,SBV+21,WAP21,SGP+21,BFF24,MPW22,XZZ+22}. However, as the detection efficiency thresholds found in~\cite{XZZ+22} are already close to the minimum values required to achieve Bell violation, it seems unlikely that our current approach would be able to outperform those results for that noise model, especially since we do not have tight bounds on $Q(\rho_{ET|00}, \rho_{ET|11})$. We hence leave a detailed analysis of that noise model for future work.

A relevant consideration for further research is the minimum block size $k$ required to yield a positive key rate with our new security condition. One drawback of the proof approach in this work, as noted in Sec.~\ref{sec:securitycond}, is that it does not provide explicit bounds on this minimum block size.
Resolving this issue would require finding a lower bound on the conditional entropy $H(C|\mathbf{ET})_{\overline \rho}$ in terms of $Q(\rho_{ET|00}, \rho_{ET|11})$ only, without any dependence on the system dimensions. The lower bound we use in our security proof is reliant on a function $g(k)$ (see Appendix~\ref{app:sufproof})
whose only known lower bound is dimension-dependent~\cite{AMV12}. To explicitly compute the minimum required block sizes in the DI setting, one would need to find a dimension-independent lower bound on $g(k)$  (possibly at the cost of changing its asymptotic scaling behaviour with respect to $k$, though such a change is not an issue for our proof), or possibly use an entirely different proof approach.

For the security conditions we derived in this work, the symmetrization bits $\mathbf{T}$ played an important role in reducing our analysis to states of the special form~\eqref{special_case_CETstate}. 
However, we note that it has been shown that if the repetition-code protocol is secure with the symmetrization procedure, then it is still secure without the symmetrization procedure~\cite{TLR20,SR08}. In particular, this implies that our sufficient security condition is still valid even in the latter setting (more precisely: if the condition~\eqref{eq:suffQ} holds for the protocol with symmetrization, then the protocol without symmetrization is secure). On the other hand, this does potentially affect the \emph{necessary} security condition --- in principle, it might be possible that there exists an attack for Eve when symmetrization is performed, but not when it is omitted. This would imply that our sufficient and necessary conditions no longer match exactly for the protocol without symmetrization. We leave for future work the question of whether this is indeed the case, as well as whether the conjecture regarding Eve's states in the necessary condition can be resolved. (On the other hand, for the modified version of the repetition-code protocol that we describe in Appendix~\ref{app:modified}, the symmetrization step indeed does not affect either the necessary or sufficient conditions, as we discuss in that appendix.)  

In this work, we have focused on the repetition-code protocol because in the device-dependent case, it is the protocol that has achieved the highest noise tolerances thus far~\cite{KL17}. However, a possible direction to investigate in future work would be other advantage distillation protocols, for instance as described in~\cite{MRD+09,arx_Myh11}. In such protocols, the NQCD might no longer be the main quantity of interest: for the repetition-code protocol, it has a structure that yields tensor-power states similar to those considered when studying the NQCD, but this may not be the case for other advantage distillation protocols.

Finally, it would be ideal if our results could be extended to the coherent attack setting. One difficulty faced in this setting is that by eliminating the IID assumption, the single-round probability distributions $\Prob(ab|xy)$ and conditional states $\sigma_{E|ab}$ (and consequently $\rho_{ET|ab}$) would no longer be well-defined without conditioning on all previous input and output values. Also, the tensor product structure of $\rho_{\mathbf{ET}|\mathbf{mm}}$ would no longer hold, in which case it unclear whether it would be possible to extend our analysis based on the NQCD (whose nature is ingrained in distinguishing tensor-power states) to this regime. 

\begin{acknowledgments}

We thank Joseph M.~Renes and Marco Tomamichel
for helpful discussions. 
Financial support for this work has been provided by the Natural Sciences and Engineering Research Council of Canada (NSERC) Alliance, and Huawei Technologies Canada Co., Ltd.
Computations were performed with the NPAHierarchy function in QETLAB~\cite{qetlab}, using the CVX package~\cite{cvxpackage,cvxbook} with solver MOSEK~\cite{mosek}.

\end{acknowledgments}

\bibliography{fullbiblio}

\newpage
\clearpage
\onecolumngrid

\appendix

\section{Proof of Theorem~\ref{theorem:1}} \label{app:sufproof}
\begin{proof}
We modify the proof of Theorem~1 in \cite{TLR20} to replace the fidelity with the NQCD. 
Our goal is to show that the Devetak--Winter condition~\eqref{eq:DWcondition} (i.e.~$H(C|\mathbf{ETM};D=1)-H(C|C';D=1)>0$) is satisfied.
To begin, we first note that conditioned on the block being accepted ($D=1$), the probability that $C\neq C'$ is
\begin{align}\label{eq:guessC'value}
\Prob(C \neq C'| D=1) = \frac{\epsilon^k}{\epsilon^k+(1-\epsilon)^k}.
\end{align}
For ease of notation, let us use $\delta_k$ to denote the right-hand-side of the above expression, i.e.
\begin{align}\label{eq:delta_k}
\delta_k \coloneqq \frac{\epsilon^k}{\epsilon^k+(1-\epsilon)^k}.
\end{align}
(Note that $\delta_k\to 0$ as $k\to\infty$.) With this notation, the term $H(C|C';D=1)$ in the Devetak--Winter condition can be written as
\begin{align}\label{eq:entBob}
H(C|C';D=1) = h\left(\delta_k\right),
\end{align}
where $h$ is the binary entropy function.

Now turning to the $H(C|\mathbf{ETM};D=1)$ term in the Devetak--Winter condition, we note that since $H(C|\mathbf{ETM};D=1) = \sum_{\mathbf{m}} \Prob(\mathbf{M}=\mathbf{m}|D=1) H(C|\mathbf{ET};\mathbf{M}=\mathbf{m} \text{ and } D=1)$, it suffices to lower-bound the entropies $H(C|\mathbf{ET};\mathbf{M}=\mathbf{m} \text{ and } D=1)$ conditioned on each value of $\mathbf{m}$. It was shown in~\cite{TLR20} (see Eq.~(S5) of the Supplemental Material for that work) that by using a continuity bound (Lemma 2 of~\cite{Win16}) to bound the contributions from cases where $C\neq C'$, the following bound holds:
\begin{align}\label{eq:contbnd}
    H(C|\mathbf{ET};\mathbf{M}=\mathbf{m} \text{ and } D=1) \geq H(C|\mathbf{ET})_{\widetilde \rho} - \delta_k - (1+\delta_k)h\left(\frac{\delta_k}{1+\delta_k}\right) ,
\end{align}
where $\widetilde \rho_{C\mathbf{ET}}$ is the state defined in~\eqref{CET_state}.

Next, we note that by applying Lemma~\ref{lemma:1} followed by a bound that relates conditional entropies to trace distance (Theorem~14 of~\cite{BH09}), we have
\begin{align}
    H(C|\mathbf{ET})_{\widetilde \rho} = H(C|\mathbf{ET})_{\overline \rho} \geq 1-d\left(\rho_{ET|00}^{\otimes k}, \rho_{ET|11}^{\otimes k}\right).
\end{align}
Hence it suffices to upper bound the trace distance in the last expression. This brings us to the core part of our proof: the characteristic property of the NQCD (as shown in Theorem~2.2 of~\cite{NS09} and Eq.~(1) of~\cite{ACM+07}) is that it is related to the trace distance $d\left(\rho_{ET|00}^{\otimes k}, \rho_{ET|11}^{\otimes k}\right)$  by
\begin{equation} \label{NQCD_asymp_prot}
\ln Q\left(\rho_{ET|00}, \rho_{ET|11}\right) = \lim_{k \rightarrow \infty} \left(\frac{1}{k}\ln \frac{1-d\left(\rho_{ET|00}^{\otimes k}, \rho_{ET|11}^{\otimes k} \right)}{2} \right).
\end{equation}
Let us define $g(k)$ to be the difference between the argument in the right-hand-side of the above expression and its limiting value (basically, $g(k)$ quantifies how fast the above limit converges):
\begin{equation} \label{g_def}
g(k) \coloneqq \frac{1}{k}\ln \frac{1-d\left(\rho_{ET|00}^{\otimes k}, \rho_{ET|11}^{\otimes k}\right)}{2} - \ln Q\left(\rho_{ET|00}, \rho_{ET|11}\right) ,
\end{equation}
in which case $\lim_{k \rightarrow \infty} g(k) = 0$ due to~\eqref{NQCD_asymp_prot}.
With this, we can write the trace distance in terms of $g(k)$ and $Q\left(\rho_{ET|00}, \rho_{ET|11}\right)$ as  $d\left(\rho_{ET|00}^{\otimes k}, \rho_{ET|11}^{\otimes k}\right) = 1 - 2 e^{k g(k)} Q\left(\rho_{ET|00}, \rho_{ET|11}\right)^k$. It follows that $Q\left(\rho_{ET|00}, \rho_{ET|11}\right)$ can be related to the conditional entropy of $\overline \rho_{C\mathbf{ET}}$ by
\begin{equation} \label{lower_bound_conditional_entropy}
    H\left(C|\mathbf{ET}\right)_{\overline \rho} \geq 2e^{k g(k)}Q\left(\rho_{ET|00}, \rho_{ET|11}\right)^k.
\end{equation}

Putting together the above inequalities and the formula~\eqref{eq:entBob}, it follows that
\begin{align}
    \frac{H(C|\mathbf{ETM};D=1)}{H(C|C';D=1)
    } \geq \frac{1}{h\left(\delta_k\right)} \left(2e^{k g(k)}Q\left(\rho_{ET|00}, \rho_{ET|11}\right)^k- \delta_k - (1+\delta_k)h\left(\frac{\delta_k}{1+\delta_k}\right) \right).
\end{align}
It was shown in \cite{TLR20} (see Eq.~(S11) in the Supplemental Material of that work) that the term $\frac{1}{h(\delta_k)}\left( -\delta_k - (1+\delta_k)h\left(\frac{\delta_k}{1+\delta_k}\right) \right)$ approaches $-1$ as $k \rightarrow \infty$. Therefore, it would be sufficient to show that the term $\frac{1}{h(\delta_k)} e^{k g(k)}Q\left(\rho_{ET|00}, \rho_{ET|11}\right)^k$ limits to some value strictly larger than $1$ as $k \to \infty$, since this would mean we have $H(C|\mathbf{ETM};D=1) \geq H(C|C';D=1)$ for sufficiently large $k$. 

Let $\beta = \frac{\epsilon}{1-\epsilon} \in [0, 1)$, so the theorem condition $Q(\rho_{ET|00}, \rho_{ET|11}) > \frac{\epsilon}{1-\epsilon}$ can be written as $Q\left(\rho_{ET|00}, \rho_{ET|11}\right)>\beta$. 
Recall that by construction $g(k)$ satisfies $\lim_{k \rightarrow \infty} g(k) = 0$, which means that for any strictly positive value $M$, we will have $g(k) \geq -M$ for all sufficiently large $k$. Specifically, if we pick $M=(\ln Q\left(\rho_{ET|00}, \rho_{ET|11}\right) - \ln \beta)/2$ (which is strictly positive since $Q\left(\rho_{ET|00}, \rho_{ET|11}\right)>\beta$), then there exists $N_1 \in \mathbb{N}$ such that for all $k > N_1$ we have
\begin{equation}
g(k) \geq - \frac{\ln Q\left(\rho_{ET|00}, \rho_{ET|11}\right) - \ln \beta}{2}.
\end{equation}
Furthermore, notice that $\delta_k \leq \beta^k$, and that there exists $N_2 \in \mathbb{N}$ such that for all $ k> N_2$, we have $\beta^k < \frac{1}{2}$. Hence for all $k>N_2$, we can upper bound $h(\delta_k)$ in terms of $\beta$ via $h(\delta_k)  
\leq 
h\left(\beta^k\right) \leq 2\beta^k \log_2\left({1}/{\beta^k}\right)$ (that last inequality follows from the fact that $h(x) \leq 2 x \log_2(1/x)$ for $x\in[0,1/2]$).
Putting these together, we see that for all $k > \max\{N_1, N_2\}$ we have
\begin{align}
    \frac{e^{k g(k)}Q\left(\rho_{ET|00}, \rho_{ET|11}\right)^k}{h(\delta_k)} &\geq \frac{e^{k g(k)} Q\left(\rho_{ET|00}, \rho_{ET|11}\right)^k}{2 k \beta^k \log_2\left(\frac{1}{\beta}\right)} \\ &= \frac{e^{k\left(g(k)+\ln Q\left(\rho_{ET|00}, \rho_{ET|11}\right) - \ln \beta \right)}}{2 k \log_2\left(\frac{1}{\beta}\right)} \\ &\geq \frac{e^{k \left({\ln Q\left(\rho_{ET|00}, \rho_{ET|11}\right) - \ln \beta}\right)/{2}}}{2 k \log_2\left(\frac{1}{\beta}\right)}.
\end{align}
Since as noted above we have $\ln Q\left(\rho_{ET|00}, \rho_{ET|11}\right) - \ln \beta > 0$ from the theorem condition $Q\left(\rho_{ET|00}, \rho_{ET|11}\right)>\beta$, it is clear that the lower bound becomes strictly larger than $1$ as $k\to\infty$, which is the desired result. (In fact, the lower bound becomes arbitrarily large as $k$ increases. This reflects the fact that for the repetition-code protocol, informally speaking we can expect that the ratio between $H(C|\mathbf{ETM};D=1)$ and $H(C|C';D=1)$ should become arbitrarily large as the block size $k$ increases, because at large block sizes the repetition-code protocol is more effective at ``distilling'' the Alice-Bob correlations over the Alice-Eve correlations.)
\end{proof}

\section{Proof of Lemma~\ref{lemma:1}}
\label{app:Uinvariance}

\begin{proof}
Let us define the unitary operator $U_X = \id_{E}
\otimes X$, where $X$ denotes the Pauli-$X$ operator. Importantly, observe that from the definition of the  $\rho_{ET|ii}$ states, the unitary $U_X$ takes $\rho_{ET|ii}$ to $\rho_{ET|\overline i \overline i}$. For 
any string $\mathbf{m} \in \mathbb{Z}_2^k$,
we define the unitary operator $U^{(\mathbf{m})} = \otimes_{i=1}^k U_X^{m_i}$. Observe that for the states $\rho_{\mathbf{ET}|\mathbf{mm}}$ as defined in~(\ref{eq:defnrhoETmm}), the unitary $U^{(\mathbf{m})}$ takes $\rho_{\mathbf{ET}|\mathbf{mm}}$ to $\rho_{ET|00}^{\otimes k}$ and $\rho_{\mathbf{ET}|\mathbf{\overline m}\mathbf{\overline m}}$ to $\rho_{ET|11}^{\otimes k}$. 

Furthermore, note that for each $\mathbf{m} \in \mathbb{Z}_2^k$, the unitary operator $\widetilde U^{(\mathbf{m})} = \id_C \otimes U^{(\mathbf{m})}$ transforms $\widetilde \rho_{C\mathbf{ET}}$ to $\overline \rho_{C\mathbf{ET}}$ (recall that the former state implicitly has a dependence on $\mathbf{m}$). Due to the unitary invariance of the von Neumann entropy, this means we have $H(C
\mathbf{ET})_{\widetilde \rho
} = H(C\mathbf{ET})_{\overline \rho}$ and $H(\mathbf{ET})_{\widetilde \rho} = H(\mathbf{ET})_{\overline \rho}$. Applying the definition of conditional von Neumann entropy, we get
\begin{equation}\label{lemma1_proof}
    H(C|\mathbf{ET})_{\widetilde \rho} = H(C\mathbf{ET})_{\widetilde \rho}-H(\mathbf{ET})_{\widetilde \rho} = H(C\mathbf{ET})_{\overline \rho}-H(\mathbf{ET})_{\overline \rho} = H(C|\mathbf{ET})_{\overline \rho}.
\end{equation}
This proves the claim.
\end{proof}

\section{Modified repetition-code protocol} \label{app:modified}

We consider a modification of the repetition-code protocol where Alice and Bob perform the same operations as specified in the main text, but the condition for accepting a block is stricter: we require that Bob's string satisfies $\widetilde{\mathbf{B}}_0 \oplus \mathbf{M} = (C', ..., C')$ for some $C' \in \{0, 1\}$ just as in the standard version, but \emph{also} that
Alice and Bob's strings $\widetilde{\mathbf{A}}_0, \widetilde{\mathbf{B}}_0$ are also both constant bit-strings (i.e.~each of them is either an all-$0$ or all-$1$ string). This would require both Alice and Bob to make a public announcement: for instance, we can say that in each block Alice announces $D_A=1$ if and only if $\widetilde{\mathbf{A}}_0$ is a constant bit-string, while Bob announces $D_B=1$ if and only if $\widetilde{\mathbf{B}}_0 \oplus \mathbf{M} = (C', ..., C')$ for some $C' \in \{0, 1\}$ \emph{and} $\widetilde{\mathbf{B}}_0$ is a constant bit-string. Alice and Bob can then set $D=1$ (i.e.~accept the block) if and only if $D_A=D_B=1$.

In the context of the security proofs, the condition for $D=1$ (i.e.~accepting the block) is now stricter, which allows us to perform some simplifications in the proofs. We begin by noting that by the same arguments as in the starting parts of the proofs in Appendices~\ref{app:sufproof} and~\ref{app:necproof}, it suffices to study the states conditioned $\mathbf{M}=\mathbf{m}$ for each possible value of $\mathbf{m}$.
However, with this modified condition for accepting the block, we see that there are only two possible values of $\mathbf{M}$ when $D=1$, namely  $\mathbf{M} = (0,\dots,0)$ and $\mathbf{M} = (1,\dots,1)$. 
Focusing first on the former case, we list the scenarios corresponding to the four possible values of $(C,C')$, conditioned on $\mathbf{M} = (0,\dots,0)$ and $D=1$ (using the fact that the relation $\mathbf{M} = \widetilde{\mathbf{A}}_0 \oplus (C,\dots,C)$ is equivalent to $\widetilde{\mathbf{A}}_0 = \mathbf{M}\oplus (C,\dots,C)$, and analogously for Bob with $\widetilde{\mathbf{B}}_0$ and $C'$):
\begin{enumerate}
\item Alice generated $C=0$ and Bob computed $C'=0$. In this case, since $\mathbf{M} = (0,\dots,0)$, Alice's string must be $\widetilde{\mathbf{A}}_0=\mathbf{M}\oplus (C,\dots,C)=(0,\dots,0)$, and since $D=1$, Bob's  
string must be $\widetilde{\mathbf{B}}_0 = \mathbf{M}\oplus (C',\dots,C')=(0,\dots,0)$. 
\item Alice generated $C=0$ and Bob computed $C'=1$. In this case, since $\mathbf{M} = (0,\dots,0)$, Alice's string must be $\widetilde{\mathbf{A}}_0=\mathbf{M}\oplus (C,\dots,C)=(0,\dots,0)$, and since $D=1$, Bob's  
string must be $\widetilde{\mathbf{B}}_0 = \mathbf{M}\oplus (C',\dots,C')=(1,\dots,1)$. 
\item Alice generated $C=1$ and Bob computed $C'=0$. In this case, since $\mathbf{M} = (0,\dots,0)$, Alice's string must be $\widetilde{\mathbf{A}}_0=\mathbf{M}\oplus (C,\dots,C)=(1,\dots,1)$, and since $D=1$, Bob's string must be $\widetilde{\mathbf{B}}_0 = \mathbf{M}\oplus (C',\dots,C')=(0,\dots,0)$. 
\item Alice generated $C=1$ and Bob computed $C'=1$. In this case, since $\mathbf{M} = (0,\dots,0)$, Alice's string must be $\widetilde{\mathbf{A}}_0=\mathbf{M}\oplus (C,\dots,C)=(1,\dots,1)$, and since $D=1$, Bob's string must be $\widetilde{\mathbf{B}}_0 = \mathbf{M}\oplus (C',\dots,C')=(1,\dots,1)$. 
\end{enumerate}

Now observe that scenarios 1--4 correspond to Eve holding conditional states $\rho_{ET|00}^{\otimes k}, \rho_{ET|01}^{\otimes k}, \rho_{ET|10}^{\otimes k}$ and $\rho_{ET|11}^{\otimes k}$ respectively, due to the values of Alice and Bob's strings in each scenario. 
Recall we want to quantify Eve's uncertainty about $C$, in the form of a lower bound on $H(C|\mathbf{ET};\mathbf{M}=(0,\dots,0) \text{ and } D=1)$. If we now informally ignore scenarios 2 and 3 as they occur with low probability for large $k$ (to see how to rigorously remove these cases, see the bound~\eqref{eq:contbnd} in Appendix~\ref{app:sufproof}, or the bound~\eqref{eq:guessC''bound} in Appendix~\ref{app:necproof} based on the condition $\rho_{ET|01}=\rho_{ET|10}$), 
we see that the state across the registers $C\mathbf{ET}$ (conditioned on $\mathbf{M}=(0,\dots,0) \text{ and } D=1$) would be precisely of the form~(\ref{special_case_CETstate}), as claimed. Hence informally speaking, Eve's task is then reduced to distinguishing between the states $\rho_{ET|00}^{\otimes k}$ and $\rho_{ET|11}^{\otimes k}$, which are $k$ copies of two different quantum states, a situation in which the NQCD naturally arises. The remainder of the analysis for this case proceeds the same way as in Appendices~\ref{app:sufproof} or~\ref{app:necproof}, although here we already have that the state on $C\mathbf{ET}$ is of the form~(\ref{special_case_CETstate}).

To finish the argument, we consider the other possible value of $\mathbf{M}$ when the block is accepted, i.e.~$\mathbf{M}=(1,\dots,1)$. Repeating the above analysis shows that the scenarios 1--4 instead correspond to Eve holding conditional states $\rho_{ET|11}^{\otimes k}, \rho_{ET|10}^{\otimes k}, \rho_{ET|01}^{\otimes k}$ and $\rho_{ET|00}^{\otimes k}$ respectively, in which case (again, after removing scenarios 2 and 3) the state across the registers $C\mathbf{ET}$ is again of the form~(\ref{special_case_CETstate}) except for a bitflip on the $C$ register. Since that bitflip does not affect the conditional entropy (in the Appendix~\ref{app:sufproof} proof) or the probability that Eve can guess $C$ (in the Appendix~\ref{app:necproof} proof), we see that the state~(\ref{special_case_CETstate}) is again basically the relevant state to consider for this case.

One interesting aspect of this modified protocol is that we do not need to use Lemma~\ref{lemma:1} in its security analysis, since by the above arguments we can focus on the state~(\ref{special_case_CETstate}) without invoking Lemma~\ref{lemma:1}.
Note that this lemma is essentially the only part of the  proofs in Appendices~\ref{app:sufproof} or~\ref{app:necproof} that relies on explicitly analyzing the symmetrization bits~$\mathbf{T}$, apart from the fact that they ensure the outcome distribution in each round has the form $\text{Pr}(01|00)=\text{Pr}(10|00)=\frac{\epsilon}{2}$, $\text{Pr}(00|00)=\text{Pr}(11|00)=\frac{1-\epsilon}{2}$. This implies that for this modified protocol, if it could be otherwise enforced that the outcome distribution takes that form without using symmetrization bits (e.g.~by a suitable analysis of the parameter-estimation step in the IID asymptotic limit), then the symmetrization step could be omitted without affecting the sufficient or necessary security conditions, apart from the minor modification of phrasing them in terms of the states $\sigma_{E|ab}$ in~\eqref{ccq_after_key_gen} rather than the states $\rho_{E|ab}$ in~\eqref{ccq_after_symm}--\eqref{eve_single_round}.

\section{Proof of Theorem~\ref{theorem:2}}
\label{app:necproof}

\begin{proof}
Our goal will be to show that the inequality~\eqref{guessC''} (i.e.~$\Prob(C \neq C'' | D=1) \leq \Prob(C \neq C'| D=1)$) holds, in which case the asymptotic keyrate cannot be positive.
First note that as observed in~(\ref{eq:guessC'value})--(\ref{eq:delta_k}), we have  $\Prob(C \neq C'| D=1) = \delta_k$.
Next, we use a result from \cite{HT22}: using the operational relation between trace distance and distinguishing probability, it was shown in that work (as Eq.~(70)) that given the condition $\rho_{ET|01}=\rho_{ET|10}$, for any message value $\mathbf{m}$ Eve can produce a bit $C''$ such that
\begin{align}\label{eq:guessC''bound}
\Prob(C \neq C'' | D=1 \text{ and } \mathbf{M}=\mathbf{m}) \leq \frac{1}{2}\left(1-\left(1-\delta_k\right) d\left(\rho_{\mathbf{ET|mm}}, \rho_{\mathbf{ET|\overline m \overline m}} \right)\right).
\end{align}
Due to the invariance of 
the trace distance under unitary transformations, the same argument as in the proof of Lemma~\ref{lemma:1} can be applied to see that $d\left(\rho_{\mathbf{ET|mm}}, \rho_{\mathbf{ET|\overline m \overline m}}\right) = d\left(\rho_{ET|00}^{\otimes k}, \rho_{ET|11}^{\otimes k}\right)$. Furthermore, the ``Fuchs--van de Graaf-type'' inequality derived as Eq.~(9) in~\cite{ACM+07} relates this to the NQCD via
$d\left(\rho_{ET|00}^{\otimes k}, \rho_{ET|11}^{\otimes k}\right) 
\geq
1 - Q\left(\rho_{ET|00}^{\otimes k}, \rho_{ET|11}^{\otimes k}\right)$. Lastly, since $\rho_{ET|00}^{\otimes k}$ and $\rho_{ET|11}^{\otimes k}$ are both tensor-power states, it can be straightforwardly shown from the definition of the NQCD that $Q\left(\rho_{ET|00}^{\otimes k}, \rho_{ET|11}^{\otimes k}\right) = Q\left(\rho_{ET|00}, \rho_{ET|11}\right)^k$
(more generally, this is a property of the NQCD that holds for all tensor-power states, and was important in proving its relevance to the topic of symmetric quantum hypothesis testing~\cite{ACM+07,ANS+08}). 
Putting together the above relations, we see that the inequality~(\ref{eq:guessC''bound}) implies
\begin{align}
\Prob(C \neq C'' | D=1 \text{ and } \mathbf{M}=\mathbf{m}) &\leq \frac{1}{2}\left(1-\left(1-\delta_k\right)\left(1- Q\left(\rho_{ET|00}, \rho_{ET|11}\right)^k \right)\right) \\
&= \frac{1}{2}\left(\delta_k + \left(1-\delta_k\right) Q\left(\rho_{ET|00}, \rho_{ET|11}\right)^k \right) . \label{eq:guessC''bound2}
\end{align}

Since the above bound holds for every $\mathbf{m}$, we have that $\Prob(C \neq C'' | D=1)$ is also upper bounded by the expression~(\ref{eq:guessC''bound2}).
Together with~(\ref{eq:guessC'value})--(\ref{eq:delta_k}), we hence see that Eve's bit $C''$ satisfies the inequality
$\Prob(C \neq C'' | D=1) \leq \Prob(C \neq C'| D=1)$ (and thus the asymptotic keyrate cannot be positive) whenever
\begin{align}\label{lemma3_proof_step_2}
\frac{1}{2}\left( \delta_k + (1-\delta_k)Q\left(\rho_{ET|00}, \rho_{ET|11}\right)^k\right) \leq \delta_k.
\end{align}
However, rearranging the above condition shows that it is in fact just equivalent to 
\begin{align}
Q\left(\rho_{ET|00}, \rho_{ET|11}\right) \leq \left( \frac{\delta_k}{1-\delta_k} \right)^{\frac{1}{k}},
\end{align}
and since 
\begin{align}
\left( \frac{\delta_k}{1-\delta_k} \right)^{\frac{1}{k}} = \left( \frac{\epsilon^k}{(1-\epsilon)^k} \right)^{\frac{1}{k}} = \frac{\epsilon}{1-\epsilon} ,
\end{align}
this is the desired result.
\end{proof}
\end{document}